\documentclass[conference]{IEEEtran}
\IEEEoverridecommandlockouts
\usepackage{cite}
\usepackage{multirow}
\usepackage{microtype}
\usepackage{longtable}
\usepackage{kantlipsum}
\usepackage{amsmath,amssymb,amsfonts}
\usepackage{algorithmic}
\usepackage[ruled,vlined]{algorithm2e}
\usepackage{graphicx}
\usepackage{tikz}
\usepackage{pgfplots}
\pgfplotsset{width=7.5cm,compat=1.12}
\usepgfplotslibrary{fillbetween}
\usetikzlibrary{bayesnet}
\usetikzlibrary{arrows}
\usepackage{color}
\usepackage{textcomp}
\usepackage{xcolor}
\def\BibTeX{{\rm B\kern-.05em{\sc i\kern-.025em b}\kern-.08em
    T\kern-.1667em\lower.7ex\hbox{E}\kern-.125emX}}
\begin{document}

\title{On Impact of Semantically  Similar Apps in Android Malware Datasets\\
{\footnotesize \textsuperscript{*}}
\thanks{Identify applicable funding agency here. If none, delete this.}
}

\author{\IEEEauthorblockN{Roopak Surendran}
\IEEEauthorblockA{\textit{Kerala,India} \\
\textit{}}

}
\maketitle

\begin{abstract}
 Malware authors reuse the same program segments found in other applications for performing the similar kind of malicious activities such as information stealing, sending SMS and so on. Hence, there may exist several semantically similar malware samples in a family/dataset. Many researchers unaware about these semantically similar apps and use their features in their ML models for evaluation. Hence, the performance measures might be seriously affected by these similar kinds of apps.  In this paper, we study the impact of semantically similar applications in the performance measures of ML based Android malware detectors. For this, we propose a novel opcode subsequence based malware clustering algorithm to identify the semantically similar malware and goodware apps. For studying the impact of semantically similar apps in the performance measures, we tested the performance of distinct ML models based on API call and permission features of malware and goodware application with/without semantically similar apps.  In our experimentation with Drebin dataset, we found that, after removing the exact duplicate apps from the dataset ($\epsilon=0$) the malware detection rate (TPR) of API call based ML models is dropped from 0.95 to 0.91 and permission based model is dropped from 0.94 to 0.90. In order to overcome this issue, we advise the research community to use our clustering algorithm to get rid of semantically similar apps before evaluating their malware detection mechanism. 
\end{abstract}


\section{Introduction}
It is known that, malware apps frequently reuse the program segments of previously detected malware apps \cite{sun2014detecting}. Also, they can add some junk codes or remove redundant codes to change the signatures. However, these malicious apps tend to preserve the malicious program segments intended for some specific functionalities such as information stealing, sending SMS and so on.   Hence, it is clear that there may exist common malicious program segments shared by the Android malware families.  
\par
Android is an open source operating system which provides specific APIs (Application Programming Interface) to perform sensitive operations such as sending SMS, making phone call and so on \cite{mcdonnell2013empirical}. For example, sendTextMessage() API call can be used for sending SMS to others. Initially, a malware author constructs a malicious program segment which is intended to perform a particular kind of malicious activity. This is done by invoking some specific API calls in a particular manner. For convenience, evolving malware apps tend to reuse these existing malicious program segments to perform the same kind of behavior. Furthermore, there exists several other existing frameworks such as kwetza for injecting the existing malicious program segments into benign applications. 
\par
Most of the existing works use machine learning algorithms for malware classification \cite{sahs2012machine}. These approaches randomly select malware and goodware samples from the dataset for training and testing the classifiers. Some of the malware or goodware apps are semantically similar and may contain similar features. These semantically similar apps can result in overrated performance of the machine learning classifier especially in holdout evaluation. So, the reported accuracies in their paper may be biased. However, the performance of the models are not highly affected in k-fold cross validation based evaluation. In this paper, we make a study about this problem and propose a clustering algorithm to filter out the semantically similar apps from malware and goodware datasets. 
\par
In recent years, many research papers have been published in the area of Android malware detection. These works are classified into static, dynamic and hybrid analysis. In static analysis, the source code level features such as API calls, permissions etc. are used for malware detection. However, in the cases of dynamic analysis, the runtime features such as system calls, network packets etc. are used. In hybrid analysis, both static and dynamic features are used. Most of the existing works use Drebin dataset for evaluating their mechanism. Drebin is a public malware dataset which contains 5560 malware apps from 179 malware families \cite{arp2014Drebin}. Because of this popularity of Drebin dataset, we have selected this Drebin dataset for studying the impact of semantically similar apps in ML models. The usage of applications with similar program segments in experimental evaluations can give biased results. So it is necessary to identify applications to similar programs segments.  For this, we propose a novel malware clustering algorithm based on opcode subsequences to filter out semantically similar apps. The researchers can use the filtered datasets for their experimental purpose for eliminating the bias in their results.  
\par
In this work, we study the impact of semantically similar apps in machine learning models for malware detection. A clustering algorithm is proposed to filter out similar applications from both goodware and malware dataset. Then, we tested the performance of ML models in various features of  malware and goodware samples with and without the semantically similar apps. We found that the performance of ML models very slightly dropped after the semantically similar apps from the dataset when k-fold cross validation technique is used. Hence, it is advised to use k-fold cross validation for evaluating the models or filter out  the semantically similar apps from malware and goodware dataset for fair evaluation. 
\begin{figure}[h!]
 \caption{Clustering of Android Apps in the Dataset}
\par
    \centering
    \includegraphics[height=2cm,width=7cm]{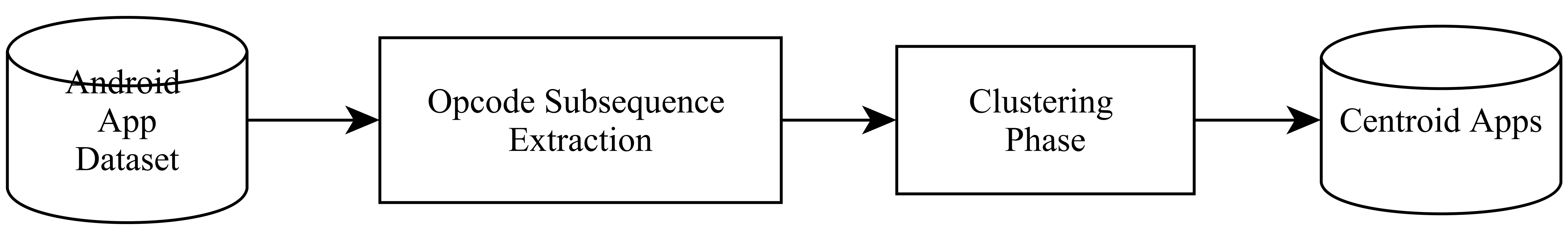}
  \end{figure}
\par    
The rest of the paper organized as follows. In Section 2, we discuss about the literature review. In Section 3, we discuss about the procedure of extracting opcode subsequences of an application. Our malware clustering algorithm is discussed in Section 4. In Section 5, we discuss about the performance of ML models in Drebin dataset with and without the semantically similar apps. In Section 6, we discuss about the limitations and future directions for our work. 

\section{Literature Review}
In this section, we discuss about the existing research works on code reuse and Android malware detection.  
\subsection{Detection of Code Reuse}
Many researchers discussed about the impact of code reuse in Android applications. By using the existing program segment dedicated for a particular functionality, an application developer can significantly save his time and effort. Moreover, it is very helpful to reduce errors or bugs in the application. However, now a days, this feature is increasingly misused by the hackers. They generate several versions of a particular kind of malware app by injecting its payload (malicious code segments) into other legitimate apps. Because of the dissimilarity in hash values, anti-malware solutions are easily evaded by these repackaged malware apps. In this section, we discuss about the main works related to the code reuse detection in malware apps.  
\par
In GroupDroid \cite{marastoni2017groupdroid}, static control flow graphs were used for clustering 4211 Android malware apps into different groups. In Droidsim \cite{sun2014detecting}, component based-control flow graphs were used for identifying similarities among malware apps and found the code reuse in a dataset of 706 malware applications. Hanna et al. \cite{hanna2012juxtapp} proposed a framework called JuxtApp for detecting code reusage among Android applications. In JuxtApp, the feature matrix of applications were constructed for measuring the similarities. JuxtApp identified the malicious code reusage in 463 vulnerable apps and 34 malware apps. In DNADroid \cite{crussell2012attack}, program dependency graphs were used for measuring the similarities among the applications and found code cloning in at least 141 apps in their dataset.  
\par
In the area of Android malware detection, many researchers considered the features of cloned apps (semantically similar) in their ML model. In this work, we study the impact of these semantically apps in ML models. For this, we propose a simple and lightweight algorithm to detect semantically similar applications in a malware dataset. Here, we used opcode subsequences as features for measuring the similarities. It is because the program segments (program statements in a function or methods) of an application can be conveniently represented with opcode subsequences. Hence, similarity/dissimilarity values can be easily computed from by comparing different sets of opcode subsequences.     
\subsection{Review on Malware Detection Mechanisms}
In existing works, machine learning algorithms are used for malware analysis because of the ability to predict malicious behavior in unseen data points \cite{liu2020review}. Most of the popular works use Drebin dataset for evaluation. Drebin is the public malware dataset which contains 5560 malware apps from 179 malware families \cite{arp2014Drebin}. Also, Drebin dataset contains malicious applications from MalGenome dataset. Hence, we have selected Drebin dataset for studying the impact of semantically similar apps in ML models.  The existing Android malware detection mechanisms use either static features such as API calls, permissions etc. or dynamic features such as system calls, network packets etc. (or the combination of both) for malware analysis. The popular static and dynamic malware analysis mechanisms in Drebin/MalGenome dataset are discussed below. 
\par
In static analysis, the features associated with the source code of an application is used for malware detection. In \cite{cen2014probabilistic},  the authors used probabilistic machine learning classifiers trained with API call based features for malware detection. In \cite{rovelli2014pmds}, the app permissions are used as input features of a machine learning classifier for malware detection. In \cite{arzt2014flowdroid}, the data flows are extracted from an application for finding malicious behavior. In \cite{xu2016iccdetector}, the intent based features are used in a machine learning classifier for malware detection. In \cite{canfora2015effectiveness}, n-gram frequencies of opcode level features are used in a machine learning classifier for malware detection.
\par
In dynamic analysis, an application is executed in an emulator or in a real device and collect the features such as system calls, network packets using the third party utilities. In \cite{kim2018runtime}, the runtime API calls are used for malware detection. In \cite{milosevic2016friend}, authors used system metric level features such as CPU, memory usages for malware detection. In \cite{canfora2015detecting} , the authors used system calls as features of supervised binary classifiers for malware detection. In \cite{zaman2015malware}, the authors used network packets as features for malware detection. 
\par
In all of the above mechanisms, the authors used entire samples in their dataset (Drebin/MalGenome) for the experimental purpose. It is known that a malware author reuses existing malicious codes to generate new varients. Hence, these malware dataset may contain several semantically similar apps.  In this paper, we study the impact of semantically similar apps in machine learning models for malware detection. We propose a clustering algorithm to filter out the semantically similar apps from datasets. After removing the semantically similar apps, we tested the performance of ML models in datasets with and without semantically similar apps. From our experimental evaluations, we conclude that the presence of semantically similar apps result in the overrated performance of ML models in malware detection. 
\section{Opcode Based Clustering Algorithm}
In this section, we investigate the impact of semantically similar apps in Android malware datasets. Our mechanism has three phases. In the first phase, we extract opcode subsequences from a set of malware and goodware applications. In the next phase, we filter out the semantically similar applications from the opcode subsequence dataset using our novel clustering algorithm. In the final phase, we evaluate the performance of ML models in dataset with and without semantically similar apps. On the basis of this performance evaluation, we will make conclusions. The clustering procedure is given in Figure 1.
\begin{figure*}[h!]
    \caption{Method Based Opcode Sub sequence of an Application}
    \centering
    \includegraphics[height=10cm,width=15cm]{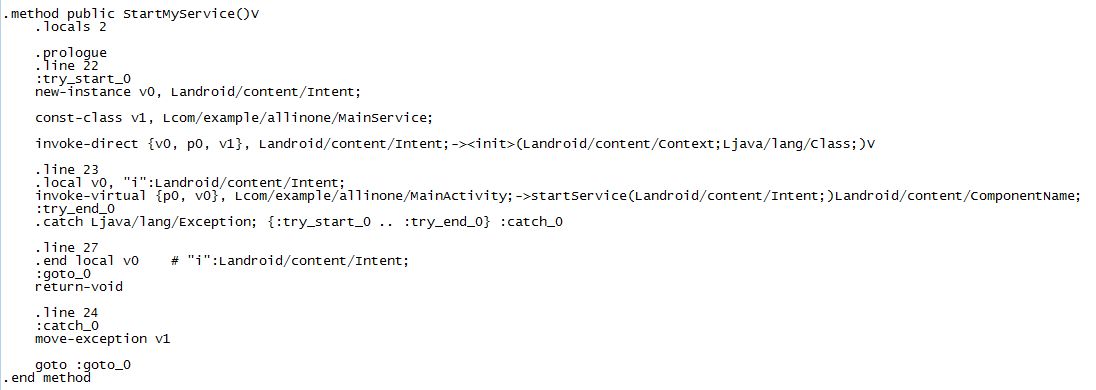}
\end{figure*}
\subsection{Extraction of Opcodes Subsequences from an Android Application}
In this section, we discuss about the procedure of extracting opcode subsequences from Android applications.
Opcodes (Operation Codes) are used to specify the kind of operations need to be performed by the device hardware.  It is a part of machine language program. The details about the opcodes in Android operating system are given in Table I.
\par
\begin{table*}[!h]
    \centering
      \caption{List Opcodes in Android Operating System}
      \resizebox{\textwidth}{!}{
    \begin{tabular}{|c|c|c|c|c|c|c|c|}
    \hline
    Hex Value & Opcode & Hex Value & Opcode & Hex Value & Opcode& Hex Value & Opcode    \\ \hline
     00&nop&01&move&02&move/from 16&03&move/16 \\ \hline
     04&move-wide/from&05&move-wide/16&07&move-object&08&move-object/from16  \\ \hline
     09&
	move-object/16&0A&
	move-result&0B&move-result-wide&0C&
	move-result-object  \\ \hline 
     0D&
	move-exception&0E&return-void&0F&return&10&
	return-wide  \\ \hline
     11 &return-object &12&
	const/4&13&
	const/16&14&const  \\ \hline
     15&const&16&const-wide/16&17&const-wide/32&18&
	const-wide  \\ \hline
     19&const-wide/high 16&1A&const-string&1B&const-string-jumbo&1C&const-class  \\ \hline
     1D&monitor-enter&1E&
	monitor-exit&1F&
	check-cast&20&
	instance-of  \\ \hline
     21&array-length&22&
	new-instance&23&
	new-array&24&filled-new-array  \\ \hline
     25&
	filled-new-array-range&26&fill-array-data&27&throw&28&goto  \\ \hline
     29&
	goto/16&2A&
	goto/32&2B&
	packed-switch&2C&sparse-switch  \\ \hline
     2D&cmpl-float&2E&
	cmpg-float&2F&
	cmpl-double&30&
	cmpg-double  \\ \hline
     31&
	cmp-long&32&
	if-eq&33&if-ne&34&
	if-lt  \\ \hline
     35&if-ge&36&if-gt&37&if-le&38&if-eqz  \\ \hline
     39&
	if-nez&3A&
	if-ltz&3B&if-gez&3C&
	if-gtz  \\ \hline
    
     3D&
	if-lez&3E&
	unused\_3E&3F&
	unused\_3F&40&
	
	unused\_40  \\ \hline
	41 & unused\_41 & 42 & unused\_42 & 43 & unused\_43 & 44 & aget  \\ \hline
	45 & 
	aget-wide & 46 & 
	aget-object & 47 & 
	aget-boolean & 48 & aget-byte \\ \hline
	49 & aget-char & 4A & 
	aget-short & 4B & 
	aput & 4C & aput-wide \\ \hline
	4D & aput-object & 4E & 
	aput-boolean & 4F & 
	aput-byte  & 50 & aput-char \\ \hline
	51 & aput-short  & 52 & iget & 53 & 
	iget-wide& 54 & 
	iget-object \\ \hline
	55 & iget-boolean & 56 & 
	iget-byte & 57 & 
	iget-char & 58 & 
	iget-short \\ \hline 
	59 & iput & 5A& 
	iput-wide & 5B & 
	iput-object & 5C & 
	iput-boolean \\ \hline
	5D & 
	iput-byte & 5E & 
	iput-char & 5F & 
	iput-short & 60 & sget \\ \hline
	  61 &sget-wide & 62 & sget-object & 63 & sget-boolean & 64 & sget-byte \\ \hline
	  65 & sget-char & 66 & sget-short & 67 & sput & 68 & 
	sput-wide \\ \hline 
	 69 & sput-object & 6A & sput-boolean & 6B & sput-byte & 6C & sput-char \\ \hline
	  6D & sput-short & 6E & 
	invoke-virtual & 6F & 
	invoke-super & 70 & 
	invoke-direct \\ \hline
	71 & invoke-static & 72 & invoke-interface & 73 & unused\_73 & 74 & 
	invoke-virtual/range \\ \hline
	75 & invoke-super/range & 76 & 
	invoke-direct/range & 77 & 
	invoke-static/range & 78 & invoke-interface-range \\ \hline
	79 & 
	unused\_79 & 7A & 
	unused\_7A & 7B & 
	neg-int & 7C & 
	not-int \\ \hline
	7D & 
	neg-long & 7E & 
	not-long & 7F & 
	neg-float & 80 & 
	neg-double \\ \hline
	81 & 
	int-to-long & 82 & 
	int-to-float & 83 & 83
	int-to-double & 84 & 
	long-to-int \\ \hline
	85 & 
	long-to-float & 86 & long-to-double & 87 & 
	float-to-int & 88 & 
	float-to-long \\ \hline
	89 & 
	float-to-double & 8A & 
	double-to-int & 8B &double-to-long  & 8C & 
	double-to-float \\ \hline
	8D & 
	int-to-byte &8E & 
	int-to-char & 8F & 
	int-to-short & 90 & 
	add-int \\ \hline
	91 & 
	sub-int & 92 & 
	mul-int & 93 & 
	div-int & 94 & 
	rem-int \\ \hline
		95 & 
	and-int & 96 & 
	or-int & 97 & 
	xor-int & 98 & 
	shl-int \\ \hline
	99 & shr-int & 9A &
	ushr-int & 9B &
	add-long& 9C& 
	sub-long \\ \hline
		9D & mul-long & 9E &
	div-long & 9F &
rem-long& A0& 
	and-long \\ \hline
		A1 & or-long & A2 &
	xor-long & A3 &
shl-long& A4& 
	shr-long \\ \hline
	A5 & ushr-long & A6 & 
	add-float  & A7 & 
	sub-float  & A8 & 
	mul-float \\ \hline 
	A9 & div-float & AA & 
	rem-float  & AB & 
	add-double  & AC & 
 sub-double \\ \hline 
 	AD & 
	mul-double & AE & 
	div-double & AF & 
	rem-double  & B0 & 
 add-int/2addr \\ \hline 
  	B1 & 
	sub-int/2addr  & B2 & 
	mul-int/2addr e & B3 & 
	div-int/2addr   & B4 & 
	rem-int/2addr  \\ \hline 
	B5 & and-int/2addr & B6 & or-int/2addr & B7 & xor-int/2addr & B8 & shl-int/2addr \\ \hline
	B9 & shr-int/2addr & BA & ushr-int/2addr & BB & 
	add-long/2addr & BC & 
	sub-long/2addr \\ \hline 
	BD & 
	mul-long/2addr & BE & 
	div-long/2addr & BF & 
	rem-long/2addr & C0 & 
	and-long/2addr \\ \hline
	C1 & or-long/2addr & C2 & xor-long/2addr & C3 & shl-long/2addr & C4 & shr-long/2addr \\ \hline
	C5 & ushr-long/2addr & C6 & add-float/2addr & C7 & sub-float/2addr & C8 & mul-float/2addr \\ \hline
	C9 & div-float/2addr & CA & rem-float/2addr & CB & 
	add-double/2addr & CC & 
	sub-double/2addr \\ \hline
	CD & 
	mul-double/2addr & CE & 
	div-double/2addr & CF & 
	rem-double/2addr & D0 & add-int/lit16 \\ \hline
	D1 & add-int/lit16 & D2 & sub-int/lit16 &D3 & mul-int/lit16 & D4 & div-int/lit16 \\ \hline
	D5 & and-int/lit16 & D6 & 
	or-int/lit16 & D7 & 
	xor-int/lit16 & D8 & 
	add-int/lit8  \\ \hline
	D9 & 
	sub-int/lit8 & DA & 
	mul-int/lit8 & DB & 
	div-int/lit8 & DC & 
	rem-int/lit8 \\ \hline    
	DD&
	and-int/lit8 & DE&
	or-int/lit8&DF&
	xor-int/lit8 & 	E0 & shl-int/lit8 \\ \hline
 E1 & shr-int/lit8 & E2 &ushr-int/lit8 & E3 & 
	unused\_E3 & E4 & unused\_E4 \\ \hline
       E5 &unused\_E5& E6 &unused\_E6& E7 & unused\_E7 & E8 & unused\_E8 \\ \hline 
         E9 &unused\_E9& EA &unused\_EA& EB &  unused\_EB & EC & unused\_EC \\ \hline 
       ED&unused\_ED&EE&execute-inline &EF&unused\_EF&F0& invoke-direct-empty \\ \hline
       F1 & unused\_F1 & F2 & iget-quick & F3 & iget-wide-quick &  F4 & 
	iget-object-quick \\ \hline
	F5 & iput-quick & F6 &iput-wide-quick &F7 & 
	iput-object-quick & F8 & 
	invoke-virtual-quick \\ \hline
	F9 & 
	invoke-virtual-quick/range & FA & invoke-super-quick & FB & invoke-super-quick/range & FC & 
	unused\_FC \\ \hline
	FD & unused\_FD & FE & unused\_FE &FF & unused\_FF & & \\ \hline
             \end{tabular}}
  
\end{table*}
In Android operating system, ART (Android Runtime) or Dalvik Virtual Machine (DVM) is responsible for handling the opcodes in the form of dex (dalvik executable format) file format \cite{brahler2010analysis}\cite{oh2012evaluation}. Android programs are written in java or kotlin language and then compiled to a `dex' (dalvik executable format) file \cite{oh2012evaluation}.  
\begin{figure*}[h!]
    \caption{Process of Extracting Opcode Sub sequences of an Application}
    \centering
    \includegraphics[height=5cm,width=12cm]{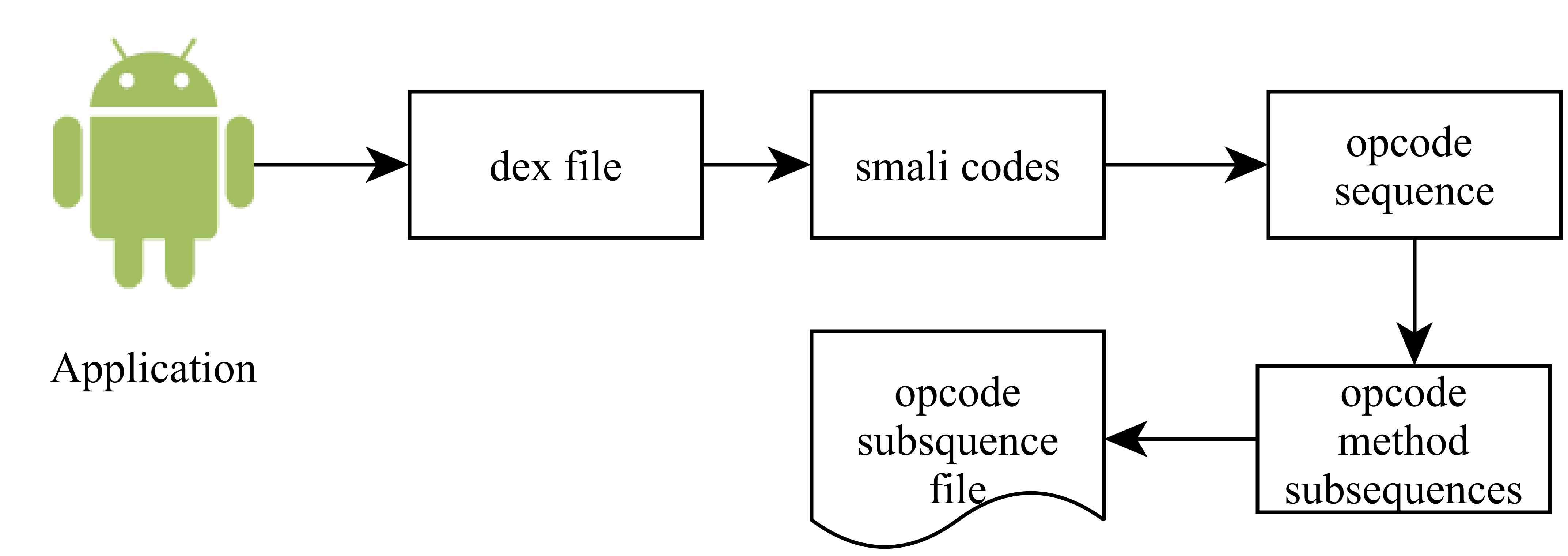}
\end{figure*}
\par
In an Android application, the program segments are written in the form of various functions or methods. Hence, there exist opcode subsequences corresponding to each program segment in that application.  We extract the set of opcode subsequences from an application and use this opcode subsequence set for representing it.  
\par
Reverse engineering tools such as apktool can be used to extract opcode sequences from the `dex' file \cite{fora2014beginners}. Apktool extracts the opcode subsequences from it. Here, we considered opcode subsequences as the sequence of opcodes in a method segment. A sample opcode subsequence of an application is given in Figure 2. The procedure of extracting opcode subsequences is given in Figure 3. 
\subsection{Clustering Opcode Subsequences}
In this section, we propose a  novel algorithm for clustering the malware apps in a
dataset. Here, we cluster the malware apps in the malware dataset by using our algorithm. Here, we use Ochiai coefficient (Euclidean distance) \cite{oguchi2020geomorphological,otsuka1936faunal,akira1957zoogeographical} for measuring the similarities between two applications. Cosine similarity works well even if the opcode subsequences of apps differs in size. Assume that $A$ and $B$ are the two sets of opcode subsequences of applications $P$ and $Q$. The Ochiai coefficient $S$ is calculated as:
\begin{equation}
    S= 1- \frac{|A \cap B|}{\sqrt{|A|\times |B|}},
\end{equation}
where $|A \cap B|$ is the number of opcode subsequences that are found in both $A$ and $B$, $|A|$ is the number of opcode subsequences in $A$ and $|B|$ is the number of opcode subsequences in $B$. 
/
/
/
/
/
/
/
/
/
/
/
/
/
/


The proposed malware clustering algorithm is based on DBSCAN algorithm \cite{xu2005survey} \cite{vassilvitskii2006k}. The algorithm accepts a malware family dataset  $X=\{X_1, X_2  X_3,…,X_n\}$ as input and gives the cluster centers $C=\{C_1,C_2,…\}$ as output. Let $\epsilon$ be the distance value ranging from 0 to 1.  The steps in our algorithm are given below.
\begin{enumerate}
\item{Initialize $j= 1$.}
\item{Select a random malware app $X_i$ from $X$ and mark $X_i$ as visited.}
\item{Find out the neighbours of $X_i$ using $\epsilon$ (All malware apps which are within the $\epsilon$ distance value are considered as neighbours).} 
\item{Form a cluster having centroid $C_j=X_i$ and update $j=j+1$.}
\begin{enumerate}
\item{Remove all the clustered apps from $X$}
\end{enumerate}
\item{Go to step 1 and repeat the process until all apps in $X$ are visited} 
\end{enumerate}
\section{Illustration of Our Clustering Algorithm in Drebin Dataset} 
In this section, we discuss about the performance of our clustering algorithm in  dataset\cite{arp2014Drebin} because of its wide acceptance and popularity in research works. Drebin dataset consists of 5560 malware applications selected from 179 malware families over a period ranging from 2010 to 2012.    
\par
Our clustering algorithm is developed and tested in an Ubuntu PC having 32 GB of memory. We reused an existing python program \cite{mclaughlin2017deep} to extract opcode subsequences of the applications in our malware family dataset. In that python code, apktool \cite{rawal2017android} is used to decompile an application and extract smali code from it. From the smali code, opcode subsequences are extracted and saved in a file. Then, we cluster all these files using our algorithm to identify the semantically similar applications.  
\par
We execute our clustering algorithm in different values of $\epsilon$. With the value $\epsilon=0$, we can remove all duplicate applications in the dataset. Also, the obtained clusters are more reliable. The number of clusters get reduced by increasing the value of $\epsilon$. Hence, by increasing the value of $\epsilon$, we can identify the highly dissimilar apps in the dataset. The  number of clusters in different $\epsilon$ values are given in Figure 4.  Here, we found that almost 50\% of apps in drebin dataset are the exact copies of others. These duplicate apps might affect the actual performance of the machine learning based malware detection mechanisms. In the next section, we investigate this with the help of API calls and permissions based classifiers.
\begin{figure}[h!]
\caption{Number of Clusters}
\centering
\begin{tikzpicture}[scale=1]
   \begin{axis}[
        xlabel=$\epsilon$,
        ylabel=$Clusters$,
        xmin=0, xmax=1,
        ymin=100, ymax=3000,
        xtick={0,0.1,0.2,0.3,0.4,0.5,0.6,0.7,0.8,0.9,1.0},
        ytick={100,600,1100,1600,2100,2600,3000}
        ]
   \addplot[smooth,mark=*,blue] plot coordinates {
        (0,2642)
       (0.1,1610)
        (0.2,1305)
        (0.3,1061)
        (0.4,891)
        (0.5,723)
        (0.6,593)
        (0.7,456)
        (0.8,321)
        (0.9,140)
        (1.0,1)
    };

      \end{axis}
    \end{tikzpicture}
\end{figure}
\begin{table*}[h!]
\caption{Distribution of Semantically Dissimilar Apps}
\centering
\begin{tabular}{|c|c|c|c|}
\hline
Dataset & $\epsilon$ & Number of Malware Samples & Number of Goodware Samples \\ \hline
Dataset 1 & 0 & 2642 & 4655 \\ \hline
Dataset 2 & 0.1 & 1650 & 3989\\ \hline
Dataset 3 & 0.2 & 1305 & 3610\\ \hline
\end{tabular}

\end{table*}
\section{Evaluation of Drebin Malware Samples with and without Semantically Similar Apps}
In this section, we make a study about the impact of semantically similar application in machine learning based malware detection mechanisms. For analyzing false positives, we collected 5500 goodware samples from Androzoo dataset \cite{allix2016androzoo}. In order to avoid the bias in the API levels, we selected the goodware samples ranging from 2010 to 2012 (same period/API level as that of drebin dataset). The overall evaluation dataset consisting of malware and goodware samples. From this dataset, we filtered out  the semantically similar goodware and malware apps and constructed  datasets with semantically dissimilar samples. The statistics of apps in the datasets are given in Table II. All the  datasets are evaluated in machine learning algorithms trained with different kinds of features. The used features are given as follows:
\begin{enumerate}
    \item{API calls;}
    \item{Permissions.}
\end{enumerate}

\begin{table*}[h!]
\caption{Selected Permissions for Malware Detection}
\centering
\begin{tabular}{|c|c|c|c|}
\hline
\textbf{SI.No} & \textbf{Permissions} &  \textbf{SI.No} & \textbf{Permissions}   \\ \hline
1 & READ\_PHONE\_STATE & 2 &WRITE\_CONTACTS  \\ \hline
3 & CALL\_PHONE &  4 & READ\_CONTACTS  \\ \hline
5 & INTERNET & 6 & SEND\_SMS  \\ \hline
7 & DISABLE\_KEYGUARD & 8 &  PROCESS\_OUTGOING\_CALLS  \\ \hline
9 & RECEIVE\_BOOT\_COMPLETED & 10 & READ\_SMS  \\ \hline
11  & FACTORY\_TEST &12 &  DEVICE\_POWER   \\ \hline
13 & HARDWARE\_TEST & 14 & CHANGE\_WIFI\_STATE \\ \hline
15  & GET\_ACCOUNTS & 16 & READ\_HISTORY\_BOOKMARKS   \\ \hline
17 &  WRITE\_APN\_SETTINGS & 18 & MODIFY\_PHONE\_STATE  \\ \hline
19  & WRITE\_HISTORY\_BOOKMARKS &  20 & ACCESS\_LOCATION  \\ \hline
21 & EXPAND\_STATUS\_BAR &22 &  WRITE\_EXTERNAL\_STORAGE  \\ \hline
23  & RECEIVE\_SMS & 24 &  WRITE\_SMS  \\ \hline
25  & ACCESS\_WIFI\_STATE & 26 & MODIFY\_AUDIO\_SETTINGS   \\ \hline
27 & ACCESS\_NETWORK\_STATE &28 &WRITE\_SETTINGS  \\ \hline
29 &READ\_EXTERNAL\_STORAGE & 30 & ACCESS\_MOCK\_LOCATION  \\ \hline
31 &USE\_CREDENTIALS & 32 & HARDWARE\_TEST  \\ \hline
33 & VIBRATE & 34& READ\_LOGS   \\ \hline
35 &CHANGE\_NETWORK\_STATE & 36 & ACCESS\_GPS  \\ \hline
37 & WAKE\_LOCK & 38 & ACCESS\_COURSE\_UPDATES  \\ \hline
39 & ACCESS\_LOCATION\_EXTRA\_COMMANDS & 40 & ACCESS\_FINE\_LOCATION  \\ \hline
41 & GET\_TASKS & 42 & RESTART\_PACKAGES \\ \hline
43 & MOUNT\_UNMOUNT\_FILESYSTEMS & 44 & INSTALL\_PACKAGES  \\ \hline
45 &   KILL \_BACKGROUND\_PROCESS  & &    \\ \hline
\end{tabular}
\end{table*}
\begin{table*}[h!]
\caption{K-Fold Cross Validation Results in Permission Classifier}
\centering
\begin{tabular}{|c|c|c|c|c|c|}
\hline
Dataset & TPR & FPR & Accuracy & Precision & F1Score \\ \hline
Overall Dataset &0.941&0.050&0.945&0.945& 0.945  \\ \hline
Dataset 1 &0.900&0.051&0.931&0.931&0.931  \\ \hline
Dataset 2 &0.855&0.054&0.920&0.920&0.920  \\ \hline
Dataset 3 &0.826&0.051&0.917&0.917&0.917  \\ \hline
\end{tabular}
\end{table*}
\subsection{Permission Analysis}
In this section, we re-implement permission based malware detection mechanism in our datasets (with and without semantically similar apps) for analyzing the impact of semantically similar in dataset. Here, we used the key permission based features mentioned in our previous work \cite{surendran2020tan}. The list of permission based features are given in Table III. We extract the permission based features of malware and goodware apps in the dataset and a Comma Separated Value (CSV) file is constructed. This CSV file is supplied to Weka framework \cite{hall2009weka} and tested in machine learning classifiers by employing 10 fold cross validation technique. We obtained a high accuracy in random forest classifier \cite{liaw2002classification}. From Table IV, we can see that the performance of the classifiers dropped very slightly in the datasets of semantically dissimilar apps. Random forest algorithm works on the basis of information gain values \cite{kent1983information}. Therefore, we have given the information gain values of permission based features in our Dataset is given in Table V. From Table V, we can see that the information gain values can be slightly affected by the semantically similar apps in the datasets. 
\begin{table*}[h!]
\caption{Changes in Information Gain Values of Permissions in the Datasets}
\centering
\begin{tabular}{|c|c|c|c|c|}
\hline
Permissions & Overall Dataset & Dataset1  & Dataset2 & Dataset3 \\ \hline
READ\_PHONE\_STATE & 0.379 & 0.334 & 0.283 & 0.264 \\ \hline
SEND\_SMS &  0.257 & 0.112 & 0.121 & 0.126 \\ \hline
RECEIVE\_BOOT\_COMPLETED & 0.189 & 0.193 & 0.143 & 0.122 \\ \hline
READ\_SMS & 0.175 &0.127 & 0.115 & 0.109 \\ \hline
RECEIVE\_SMS & 0.160 & 0.068 & 0.088 & 0.096 \\ \hline
ACCESS\_WIFI\_STATE & 0.138 & 0.209 & 0.169 & 0.142 \\ \hline
WRITE\_EXTERNAL\_STORAGE & 0.120 & 0.114 & 0.108 & 0.101 \\ \hline
WRITE\_SMS & 0.096 & 0.083 & 0.074 & 0.070 \\ \hline
WAKE\_LOCK & 0.081 & 0.073 & 0.057 & 0.049 \\ \hline
INTERNET & 0.079 & 0.060 & 0.052 & 0.053 \\ \hline
\end{tabular}
\end{table*}

\subsection{API Call Analysis}
In this section, we re-implement API call based malware detection mechanism in our datasets  for analyzing the impact of semantically similar in dataset. Here, we reused the key API call based features mentioned in our previous work \cite{surendran2020tan}. The list of API call based features are given in Table VI. We extract the API call based features of malware and goodware apps in the dataset and a Comma Separated Value (CSV) file is constructed. This CSV file is supplied to Weka framework and tested in machine learning classifiers by employing 10 fold cross validation technique. We obtained a high accuracy in random forest classifier. From Table VII, we can see that the performance of the classifiers dropped very slightly in the datasets of semantically dissimilar apps. Random forest algorithm works on the basis of information gain values. Therefore, we have given the information gain values of API Call based features in our Dataset is given in Table VIII. From Table VIII, we can see that the information gain values can be slightly affected by the semantically similar apps in the datasets. 
\begin{table*}[h!]
\caption{Selected API Calls for Malware Detection}
\centering
\begin{tabular}{|c|c|c|c|}
\hline
\textbf{SI.No} & \textbf{API Calls} & \textbf{SI.No} & \textbf{API Calls}  \\ \hline
1 & getNetworkType & 18 & getDisplayMessageBody \\ \hline
2 &getNetworkOperator&19 &getPackageInfo \\ \hline
3 &loadClass&20&getLastKnownLocation\\ \hline
4 &getMessage& 21 &getAppPackageName \\ \hline
5 &getMethod&22 &getCookies \\ \hline
6 &getClassLoader&23 &isProviderEnabled \\ \hline
7 &GetLongitude&24&getSimOperatorName \\ \hline
8 &GetLatitude& 25 &getDeviceId \\ \hline
9 &createFromPdu&26 &getCertStatus \\ \hline
10 &getInputStream&27&getSimSerialNumber \\ \hline
11 &getOutputStream&28&getLine1Number \\ \hline
12 &getWifiState& 29& killProcess \\ \hline
13&abortBroadCast& 30 & exec \\ \hline
14 &RequestFocus&31& getAppPackageName \\ \hline
15&getSubscriberId&32 &setSerialNumber \\ \hline
16&getDisplayOriginatingAddress &33 & getSessions  \\ \hline
17&sendTextMessage&34 &getCredential \\ \hline
\end{tabular}
\end{table*}
\begin{table*}[h!]
\caption{K-Fold Cross Validation Results in API Call Classifier}
\centering
\begin{tabular}{|c|c|c|c|c|c|}
\hline
Dataset & TPR & FPR & Accuracy & Precision & F1Score \\ \hline
Overall Dataset &0.954&0.046&0.957&0.957& 0.957 \\ \hline
Dataset 1 &0.91&0.037&0.944&0.944&0.944  \\ \hline
Dataset 2 &0.856&0.044&0.929&0.929& 0.928  \\ \hline
Dataset 3 &0.831&0.044&0.925&0.925&0.924  \\ \hline
\end{tabular}
\end{table*}
\begin{table*}[h!]
\caption{Changes in Information Gain Values of API Calls in the Datasets}
\centering
\begin{tabular}{|c|c|c|c|c|}
\hline
API Call & Overall Dataset & Dataset1  & Dataset2 & Dataset3 \\ \hline
getDeviceId & 0.205 & 0.301 & 0.249 & 0.229 \\ \hline
sendTextMessage &  0.175 & 0.110 & 0.106 & 0.108 \\ \hline 
getLine1Number &  0.164 & 0.224 & 0.194 & 0.180 \\ \hline
getNetworkOperator &  0.157 & 0.128 & 0.087 & 0.069 \\ \hline 
getSubscriberId &  0.154 & 0.199 & 0.228 & 0.237 \\ \hline
createFromPdu &  0.082  & 0.157 & 0.062 & 0.065 \\ \hline
abortBroadcast &  0.081 & 0.047 & 0.055 & 0.059 \\ \hline
getSimOperatorName &  0.068 & 0.093 & 0.055 &  0.045 \\ \hline 
getSimSerialNumber & 0.065 & 0.114 & 0.128 & 0.122 \\ \hline
 getCellLocation &  0.049 & 0.076 & 0.064 & 0.059 \\ \hline
\end{tabular}
\end{table*}
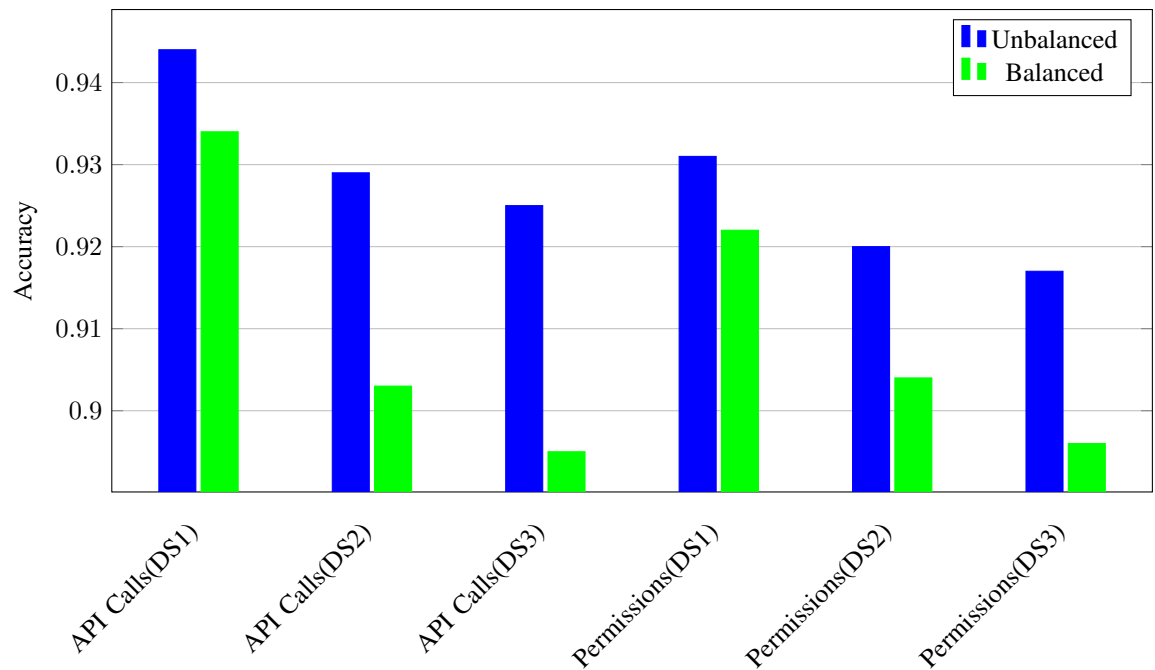
\begin{figure*}
\centering 
\begin{tikzpicture}
    \begin{axis}[
        width  = 0.85*\textwidth,
        height = 8cm,
        major x tick style = transparent,
        ybar,
        bar width=14pt,
        ymajorgrids = true,
        ylabel = {Accuracy},
        symbolic x coords={API Calls(DS1),API Calls(DS2),API Calls(DS3),Permissions(DS1),Permissions(DS2),Permissions(DS3)},
        xtick = data,
        scaled y ticks = false,
         xtick=data,
x tick label style={rotate=45, anchor=east, xshift=-1.5mm, yshift=-2mm},
    ]
        \addplot[style={blue,fill=blue,mark=none}]
            coordinates {(API Calls(DS1), 0.944) (API Calls(DS2),0.929) (API Calls(DS3),0.925) (Permissions(DS1), 0.931) (Permissions(DS2),0.920) (Permissions(DS3),0.917)};

        \addplot[style={green,fill=green,mark=none}]
            coordinates {(API Calls(DS1),0.934) (API Calls(DS2),0.903) (API Calls(DS3),0.895) (Permissions(DS1), 0.922) (Permissions(DS2),0.904) (Permissions(DS3),0.896)};

        \legend{Unbalanced, Balanced}
    \end{axis}
\end{tikzpicture}
\caption{Accuracy of Balanced Datasets without Semantically Similar Apps }
\end{figure*}
\section{Testing the Performance in Balanced Datasets}
In this section, we evaluate the performance of API call and permission based classifiers in balanced datasets. From Table II, we can see that the number of unique apps in goodware dataset is higher than that of malware dataset. That is, the distribution of apps in the classifier is not unique and class imbalance problem may occur in evaluation. Hence, it is required to evaluate the performances in balanced datasets before confirming our findings. From the goodware datasets (dataset 1 , dataset 2 and dataset 3), we removed some random goodware apps for balancing the dataset. After balancing the datasets, we evaluated the performances in API call and permission based classifiers. The performance of API call and permission based classifiers in both balanced and unbalanced datasets are given in Fig 8. From Fig 8, we can see that the performance of the classifier is dropped after balancing the dataset. 
\section{Overrated Performance in Holdout Evaluation}
In this section, we illustrate the performance bias due to the semantically similar apps in test dataset. In ML based Android malware detection, a malware researcher randomly divides the dataset samples to train and test set for evaluation. Most of this time, he unaware about the duplicate copies in the dataset. The rate of duplicate samples in the test dataset may significantly affect the performance of the model. So, it is very difficult to generalize the model in accurately detecting the diverse malware apps. Here, we illustrate this phenomenon in API call based classifier.
\par
We trained our API call and permission based classifier with the features diverse malware and goodware samples and tested with more duplicate malware and goodware samples. Further, the test dataset is constructed with more semantically similar apps of more malware samples those have more malicious features and more semantically similar apps of more goodware samples those have very few malicious features. Here,we followed thumb rule (80:20) for train-test split. Also, we shuffled the train and test set samples until obtaining the accuracy of 1.  The performance metrics are given in Table IX. From the Table IX, it is clear that it is possible for a researcher to report his desired performance by shuffling the datasets in holdout evaluation. So, it has been advised to use our clustering algorithm to remove semantically similar apps from the dataset before holdout evaluation. 
\begin{table*}[h!]
\caption{Holdout Evaluation Results in API Call and Permission Classifier}
\centering
\begin{tabular}{|c|c|c|c|c|c|}
\hline
Classifier & TPR & FPR & Accuracy & Precision & F1Score \\ \hline
API Call Classifier &1&0&1&1&1  \\ \hline
Permission Classifier &1&0&1&1&1  \\ \hline
\end{tabular}
\end{table*}
\par
\par
\par
\par
\par
\section{Discussion and Conclusions}
In this work, we proposed a clustering mechanism to assess the impact of semantically similar apps in Android malware dataset. We found that, the presence of semantically similar apps especially duplicate apps those influence the performance of ML models in hold out evaluation. So, it has been advised to filter out all semantically similar apps before performing the holdout evaluation. 
\par
Our clustering algorithm have some limitations which affects the clustering process.  In opcode injection attack, it is possible for an adversary to inject irrelevant opcodes in between the opcode subsequences of an application \cite{zhang2020andropgan}. In such cases, the application has not become the part of any cluster. In future, we will explore some other additional features such as API calls and permission sequences for efficient clustering these apps. 
\par
In our experiments, the decompilation errors has occurred in some applications. Due to this decompilation errors, we cannot cluster these apps. In future, we will investigate the reason behind this decompilation errors and design some new tools to decompile these apps. 

\section*{Acknowledgment}
\bibliography{reference}
\bibliographystyle{ieeetr}

\end{document}